## Solution calorimetry investigations of new phase $BaCe_{0.6}Y_{0.3}In_{0.1}O_{2.8}$

N.I. Matskevich, Th. Wolf, M.Yu. Matskevich, A.N. Bryzgalova, T.I. Chupakhina,
I.V. Vyazovkin

The preparation of $BaCeO_3$ doped by yttrium and indium oxides ($BaCe_{0.6}Y_{0.3}In_{0.1}O_{2.8}$) has been performed by solid-state reaction from $BaCO_3$, $CeO_2$, $Y_2O_3$, $In_2O_3$. The compound $BaCe_{0.6}Y_{0.3}In_{0.1}O_{2.8}$ has been synthesized for the first time. The X-ray measurements have showed that $BaCe_{0.6}Y_{0.3}In_{0.1}O_{2.8}$ has an orthorhombic structure (space group *Pmcn*). The standard formation enthalpies of $BaCe_{0.6}Y_{0.3}In_{0.1}O_{2.8}$ have been determined by solution calorimetry combining the solution enthalpies of $BaCe_{0.6}Y_{0.3}In_{0.1}O_{2.8}$ and $BaCl_2$ + $0.6CeCl_3$ + $0.3YCl_3$ + $0.1InCl_3$ mixtures in 1 M HCl with 0.1 M KI at 298.15 K and literature data. It has been obtained that above-mentioned mixed oxide is thermodynamically stable with respect to their decomposition into binary oxides at room temperatures. It has been also shown that $BaCe_{0.6}Y_{0.3}In_{0.1}O_{2.8}$ has been more thermodynamically favored than $BaCe_{0.9}In_{0.1}O_{2.95}$.

### 1. Introduction

Compounds on the basis of alkali-earth metal cerates (mainly, barium and strontium) with perovskite structure are perspective materials to be used in fuel cells as electrolytes, electrocatalysis, selective ceramic membranes, etc. They have sufficient high ionic conductivity at high temperatures [1-10]. The proton conductivity of the phases is about 0.1 S $cm^{-1}$ at 900°C. For instance, barium cerate exhibits protonic conduction in moist atmospheres when it is partially substituted by trivalent dopant cation, e.g., rare-earth elements, for cerium, and is counted as a perovskite-type high-temperature proton conductor (HTPC). Since the proton conductivity of proton conductors has low temperature dependence, there is a relatively high conductivity at low temperatures; for instance, $BaCe_{0.9}Y_{0.1}O_{2.95}$ has a conductivity a little lower than $10^{-2}$ S $cm^{-1}$ at 400°C. The material may thus be useful for low-intermediate temperature SOFC operating at 400–600°C as well. Only cerates doped by trivalent cations have a high ionic conductivity.

Here, it is necessary to note that the dopant solubility limit for rare-earth elements is less than 20% of the available B sites, and therefore, the possibility of introducing protonic defects into the matrix is also limited [4, 6]. But it is important to increase the dopant solubility for further practical applications since it allows one to vary smoothly the proton conductivity in large interval. It was shown in papers [11-14] that it is possible to increase the dopant solubility limit in barium cerate by introducing indium instead of cerium. However, the substitution on indium leads to decreasing the proton conductivity in comparison with the same content of rare-earth elements. For this reason the authors [4, 6] synthesized two new barium cerate phases doped simultaneously by indium and yttrium with the following compositions: $BaCe_{0.7}In_{0.2}Y_{0.1}O_{2.85}$, $BaCe_{0.7}In_{0.1}Y_{0.2}O_{2.85}$.

A lot of investigations were devoted to the protonic conductivity of doped cerates. However, it is not sufficient in practical applications that materials have a high protonic conductivity. It is necessary to study other properties as well. Thermodynamic stability is one of the important properties of doped cerates. The thermodynamic stability of complex oxides may have an impact on the mechanical stability of the microstructure of corresponding ceramics.

Our paper represents the synthesis of new phase $BaCe_{0.6}Y_{0.3}In_{0.1}O_{2.8}$, its cell parameters and thermodynamic stability with respect to binary oxides (BaO, $Y_2O_3$, $In_2O_3$, $CeO_2$). Dissolution enthalpy and standard formation enthalpy have been obtained as well.

## 2. Experimental Section

### *Preparation of samples*

In the present study we report the preparation of new $BaCeO_3$ doped by indium and yttrium oxides ($BaCe_{0.6}Y_{0.3}In_{0.1}O_{2.8}$) and their thermochemical properties. In comparison to papers [4, 6] we increased the doping by yttrium up to 0.3. It was possible to increase the dopant limit of yttrium because we added indium and yttrium together. Here we would like to note the following. Ionic radiuses of cerium, yttrium, and indium according to Shannon's paper [15] are the following: 0.087 nm, 0.090 nm, 0.080 nm. Criterion of structure stability is tolerance factor which for perovskite structure varies from 0.8 up to 1. Stability is increasing with increasing tolerance factor. Tolerance factors for $BaCeO_3$ and $BaCe_{0.8}Y_{0.2}O_{2.9}$ are 0.94 and 0.938 correspondently. Tolerance factor for $BaCe_{0.6}Y_{0.3}In_{0.1}O_{2.8}$ is 0.939. It gave basis to think that phase $BaCe_{0.6}Y_{0.3}In_{0.1}O_{2.8}$ can be prepared as individual. Below we will confirm it.

Polycrystalline sample of $BaCe_{0.6}Y_{0.3}In_{0.1}O_{2.8}$ was prepared by solid state synthesis. Barium carbonate (CERAC, TM incorporated, 99.999% pure, USA), cerium oxide (99.9%, Vetron GMBH, Karlsruhe), yttrium oxide (Venton, Alfa Produkte, 99.9999 %) and indium oxide (Johnson Matthey, 99.99%) were used as starting compounds. All starting materials were annealed at high temperature before solid state reactions. $Y_2O_3$, $In_2O_3$ and $CeO_2$ were annealed at 1023 K in air for 10 h. $BaCO_3$ was annealed at 650 K in air for 4 h.

The powders of $BaCO_3$, $CeO_2$, $Y_2O_3$ and $In_2O_3$ were mixed by ball milling in an agate container with agate balls using a planetary mill (FRITSCH pulverisette) during 72 h. The ground materials were palletized using a 10 mm diameter die and fired at 1300 K for 70 h, 1400 K for 10 h, 1700 K for 24 h using CARBOLITE furnace.

Anhydrous $BaCl_2$, $CeCl_3$, $InCl_3$ and $YCl_3$ were prepared as it was described in papers [11, 13]. All manipulations with $CeCl_3$, $BaCl_2$, $YCl_3$ and $InCl_3$ were performed in a dry box (pure Ar gas).

### *Thermochemical experiments*

The solution calorimetric experiments were carried out in an automatic calorimeter with an isothermal jacket. The calorimeter consists of a Dewar vessel with a brass cover (V = 200 ml). The platinum resistance thermometer, calibration heater, cooler, mixer, and device to break the ampoules were mounted on the lid closing the Dewar vessel. The construction of the solution calorimeter and the experimental procedure were described elsewhere [16-19]. The resistance of platinum resistance thermometer was measured by high precision voltmeter Solartron 7061. The voltmeter was connected with the computer through interface and the program written in Matlab at our laboratory. The program allows one to measure and record the temperature of vessel, calibrate the instrument with precise injections of electrical energy and calculate calorimeter constants and enthalpies. The calorimetric vessel was maintained at 298.15 K. Dissolution of potassium chloride in water was performed to calibrate the calorimeter. The obtained dissolution heat of KCl was 17.41 ± 0.08 kJ $mol^{-1}$ (the molality of the final solution was 0.028 mol $kg^{-1}$, T = 298.15 K). The literature data are: 17.42 ± 0.02 kJ $mol^{-1}$) [20], 17.47 ± 0.07 kJ $mol^{-1}$[21].

The amount of substance used ($BaCe_{0.6}Y_{0.3}In_{0.1}O_{2.8}$ ) was about 0.08 g. All compounds were stored in a dry box to prevent interaction with moisture or $CO_2$.

The derivation of the enthalpy of formation of $BaCe_{0.6}Y_{0.3}In_{0.1}O_{2.8}$ was done using the following scheme of thermochemical reactions (see, Table 1). The principal scheme is based on the dissolution of barium cerate doped by indium and neodymium oxides as well as the mixture of barium chloride, cerium chloride, indium chloride and yttrium chloride in hydrochloric acid [HCl (sol)] with KI. The molar concentration of metal chlorides was the same as in paper [22]. A mixture of $BaCl_2$, $CeCl_3$, $InCl_3$, $YCl_3$ was prepared in ratio 1: 0.6:0.1:0.3.

## 3. Results and Discussion

Phase purity and identity of $BaCe_{0.6}Y_{0.3}In_{0.1}O_{2.8}$ was confirmed by X-ray powder diffraction (XRD) using a STADI-P, Stoe diffractometer, Germany (Mo radiation) and ARL ADVANT'XP sequential X-ray Fluorescence Spectrometer. The sample was shown to be phase-pure ceramics with an orthorhombic structure (space group *Pmcn*). The refined cell parameters obtained for $BaCe_{0.6}Y_{0.3}In_{0.1}O_{2.8}$ : a = 0.6152(1) nm, b = 0.8926(1) nm, c = 0.6148(1)nm. Powder X-ray diffraction pattern of barium cerate doped by indium and yttrium oxides was presented in Fig. 1.

The measured enthalpies of solution of $BaCe_{0.6}Y_{0.3}In_{0.1}O_{2.8}$ and $BaCl_2$ + $0.6CeCl_3$ + $0.3YCl_3$ + $0.1InCl_3$ were determined as: $\Delta_{sol}H^o{}_1$(298.15 K) = –349.88 ± 4.33 kJ/mol (n = 5), $\Delta_{sol}H^o{}_2$(298.15 K) = –157.64 ± 1.31 kJ/mol (n = 6). Errors were calculated for the 95% confidence interval using standard procedure of treatment of experimental data [16-19].

The measured enthalpies of dissolution were used for calculating the enthalpy of the reaction

Ba (s) +0.1Y (s) + 0.3Y(s) + 0.6Ce (s) + $1.4O_2$ (g) = $BaCe_{0.6}\ In_{0.1}Y_{0.3}O_{2.8}$ (s) (1)

according to the equation

$$\Delta_rH_{14}{}^o = -\Delta_{sol}H_1{}^o + \Delta_{sol}H_2{}^o + \Delta_{sol}H_3{}^o - \Delta_{sol}H_4{}^o - \Delta_{sol}H_5{}^o + \Delta_{sol}H_6{}^o + \Delta_{sol}H_7{}^o + \Delta_{sol}H_8{}^o - \Delta_{sol}H_9{}^o + \Delta_{sol}H_{10}{}^o + \Delta_{sol}H_{11}{}^o + \Delta_{sol}H_{12}{}^o + \Delta_{sol}H_{13}{}^o.$$

Here, $\Delta_rH_{14}{}^o$ = $\Delta_fH^o$($BaCe_{0.6}\ In_{0.1}Y_{0.3}O_{2.8}$, s, 298.15 K) = –1597.30 ± 5.02 kJ/mol is the standard formation enthalpy of barium cerate doped by neodymium oxide.

To calculate this value we used experimental data measured by us and literature data for formation enthalpies of different compounds and processes taken from Ref. [20, 22] and presented in Table 1.

The data for formation enthalpies of BaO, $CeO_2$, $Y_2O_3$ and $In_2O_3$ taken from Ref. [20] were used to calculate the enthalpies of formation of $BaCe_{0.6}\ In_{0.1}Y_{0.3}O_{2.8}$ from binary oxides as following:

BaO(s) + $0.6CeO_2$(s) + $0.15Y_2O_3$(s) + $0.05In_2O_3$(s) = $BaCe_{0.6}Y_{0.3}In_{0.1}O_{2.8}$ (s) (2)

$\Delta_{ox}H^o$ = –63.1 ± 5.0 kJ/mol

All the data were reported for the first time.

To understand whether the $BaCe_{0.6}Y_{0.3}In_{0.1}O_{2.8}$ phase is stable or unstable with respect to decomposition to BaO(s) + $0.6CeO_2$(s) + $0.15Y_2O_3$(s) + $0.05In_2O_3$(s) mixture it is necessary to know the Gibbs energies ($\Delta G = \Delta H$ - T$\Delta S$). There is no entropy of $BaCe_{0.6}Y_{0.3}In_{0.1}O_{2.8}$ phase in literature. This value was estimated using entropies of $Y_2O_3$, $In_2O_3$, $BaCeO_3$, $CeO_2$ taken from references [20, 22]. Using the formation enthalpy of reaction (2), the Gibbs energy for the process (2) was estimated correspondently: $\Delta_{ox}G^{o}$ (298.15 K) = –66.1 ± 5.0 kJ/mol

As can be seen, above-mentioned mixed oxide is thermodynamically stable with respect to their decomposition into binary oxides at room temperatures. It is not obvious result for this class of compounds because there is a discussion about thermodynamic stability of $BaCeO_3$ [23]. S. Gopalan and A.V. Virkar predicted from galvanic measurements that $BaCeO_3$ is unstable below 1090 °C in the presence of 1 atm $CO_2$ [23]. For the reaction BaO + $CeO_2$ = $BaCeO_3$ they gave the values for standard enthalpy and entropy as 95.3 kJ/mol and 105.402 J/mol K respectively. That means that barium cerate is thermodynamically unstable with respect to its oxides below 631 °C, although it is kinetically stable.

Then, it is interesting to compare the thermodynamic stability of $BaCe_{0.6}Y_{0.3}In_{0.1}O_{2.8}$ and $BaCe_{0.9}In_{0.1}O_{2.95}$ which was measured earlier by us [24]. The formation enthalpy of $BaCe_{0.9}In_{0.1}O_{2.95}$ from oxide is $\Delta_{ox}H^{o}$ = –36.2 ± 3.4 kJ/mol. So, it is possible to see that $BaCe_{0.6}Y_{0.3}In_{0.1}O_{2.8}$ is more thermodynamically stable than $BaCe_{0.9}In_{0.1}O_{2.95}$ at room temperature. It is also possible to write the following reaction:

$$2BaCe_{0.6}Y_{0.3}In_{0.1}O_{2.8} = BaCe_{0.9}In_{0.1}O_{2.95} + 0.3Y_2O_3 + 0.7BaO + 0.05In_2O_3 + 0.3BaCeO_3 \qquad (3)$$

Using data of this paper on $\Delta_{ox}H^{o}$ ($BaCe_{0.6}Y_{0.3}In_{0.1}O_{2.8}$, 298.15 K) = –63.1 ± 5.0 kJ/mol, the enthalpy of $BaCe_{0.9}In_{0.1}O_{2.95}$ from binary oxides ($\Delta_{ox}H^{o}$ = –36.2 ± 3.4 kJ/mol) and the enthalpy of $BaCeO_3$ from binary oxides taken from [22] it is possible to estimate the enthalpy of reaction (3) as following: $\Delta_{r}H^{o}$ = +74.5 kJ/mol. So, $BaCe_{0.6}Y_{0.3}In_{0.1}O_{2.8}$ is thermodynamically stable with respect to decomposition to mixture $BaCe_{0.9}In_{0.1}O_{2.95}$,$Y_2O_3$, BaO, $In_2O_3$, $BaCeO_3$.

## Conclusions

In this paper for the first time we synthesized the compound $BaCe_{0.6}Y_{0.3}In_{0.1}O_{2.8}$ by solid-state reaction. Compound has an orthorhombic structure (space group *Pmcn*). We also measured the standard formation enthalpies of $BaCe_{0.6}Y_{0.3}In_{0.1}O_{2.8}$ by solution calorimetry in 1 M HCl with 0.1 M KI. We determined the stability of Y-doped barium cerates with respect to mixtures of binary oxides. On the basis of these data we established that above-mentioned mixed oxide is thermodynamically stable with respect to their decomposition into binary oxides at room temperatures. $BaCe_{0.6}Y_{0.3}In_{0.1}O_{2.8}$ is also more thermodynamically favoured than $BaCe_{0.9}In_{0.1}O_{2.95}$.

## Acknowledgments

This work is supported by Karlsruhe Institute of Technology (Germany), Russian Fund of Basic Research (Project No 12-08-31556) and Program of Fundamental Investigation of Siberian Branch of the Russian Academy of Sciences.

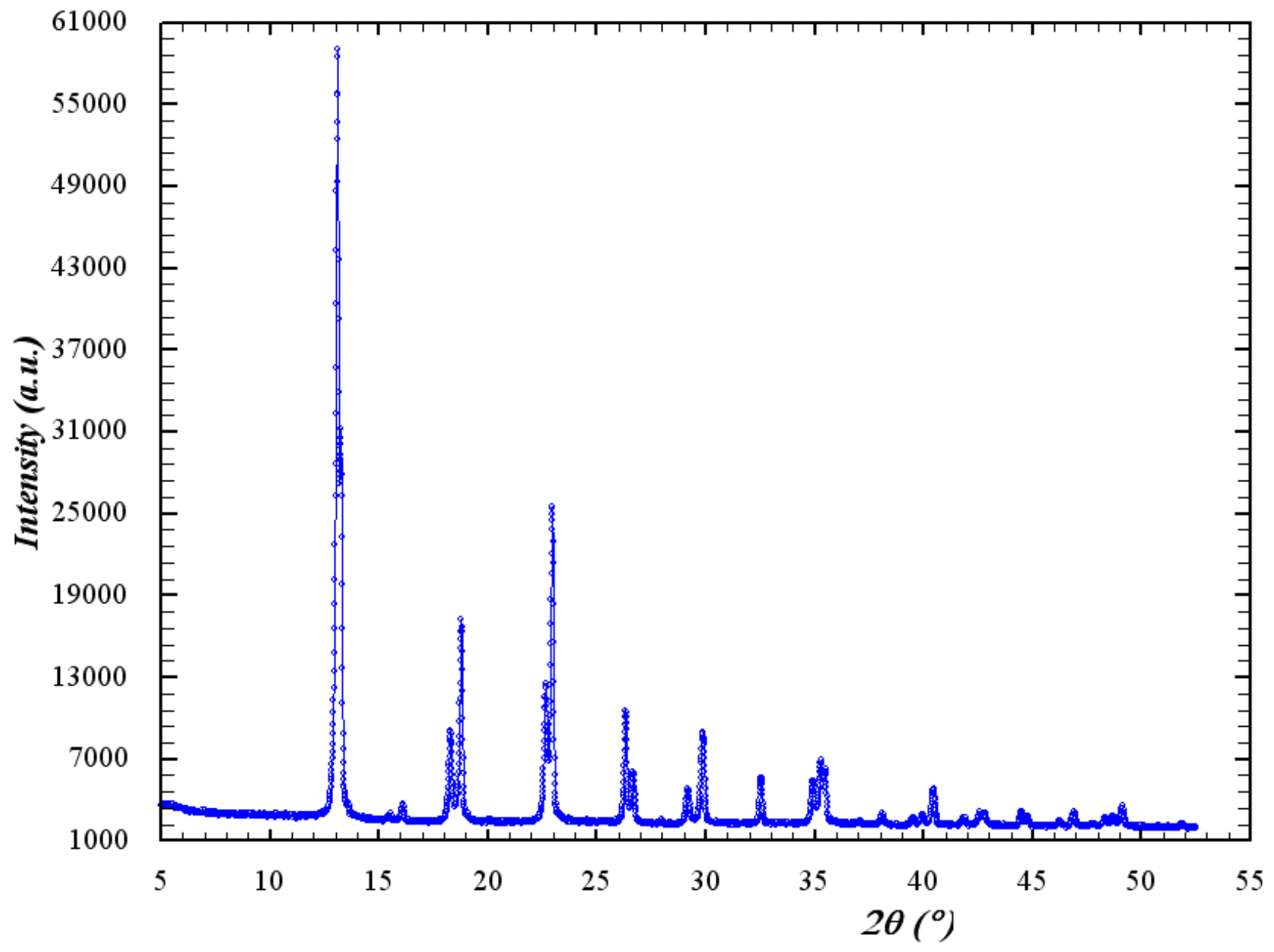

Fig.1. Experimental X-ray diffractogram of $BaCe_{0.6}Y_{0.3}In_{0.1}O_{2.8}$

Table 2. Thermochemical cycle for the determination of the enthalpy of formation of $BaCe_{0.6}Y_{0.3}In_{0.1}O_{2.8}$

| Reactions | $\Delta_{sol}H^{o}_{m}$, kJ | Ref. |
| --- | --- | --- |
| 1) $BaCe_{0.6}Y_{0.3}In_{0.1}O_{2.8}$(s) + (5.6HCl + 0.9KI)(sol) = ($BaCl_2$ + 0.6$CeCl_3$ + 0.3$YCl_3$ + 0.1$InCl_3$ + 0.6KCl + 0.3$KI_3$ + 2.8$H_2O$)(sol) | –349.88 ± 4.33 | This work |
| 2) $BaCl_2$(s) + 0.6$CeCl_3$(s) + 0.3$YCl_3$(s) + 0.1$InCl_3$ (s) + (solution 1) = ($BaCl_2$ + 0.6$CeCl_3$ + 0.3$YCl_3$ + 0.1$InCl_3$)(sol) | –157.64 ± 1.31 | This work |
| 3) 2.8$H_2$(g) + 1.4$O_2$(g) + (solution 1) = 2.8$H_2O$(sol) | –800.36 ± 0.13 | [22] |
| 4) 0.9KI(s) + (solution 1) = 0.9KI(sol) | +18.75 ± 0.39 | [22] |
| 5) 0.9K(s) + 0.45$I_2$(s) = 0.9KI(s) | –296.24 ± 0.15 | [22] |
| 6) 0.3K(s) + 0.45$I_2$(s) + (solution 1) = 0.3$KI_3$(sol) | –90.82 ± 0.10 | [22] |
| 7) 0.6KCl(s) + (solution 1) = 0.6KCl(sol) | +10.81 ± 0.03 | [22] |
| 8) 0.6K(s) + 0.3$Cl_2$(g) = 0.6KCl(s) | –261.88 ± 0.09 | [22] |
| 9) 2.8$H_2$(g) + 2.8$Cl_2$(g) + (solution 1) = 5.6HCl(sol) | –920.41 ± 0.06 | [22] |
| 10) Ba(s) + $Cl_2$(g) = $BaCl_2$(s) | –855.15 ± 1.73 | [22] |
| 11) 0.6Ce(s) + 0.9$Cl_2$(g) = 0.6$CeCl_3$(s) | –636.33 ± 0.45 | [22] |
| 12) 0.1In (s) + 0.15$Cl_2$(g) = 0.1$InCl_3$(s) | –53.72 ± 0.84 | [20] |
| 13) 0.3Y(s) + 0.45$Cl_2$(g) = 0.3$YCl_3$(s) | –299.99 ± 0.82 | [20] |
| 14) Ba (s) +0.1Y (s) + 0.3Y(s) + 0.6Ce (s) + 1.4$O_2$ (g) = $BaCe_{0.6}In_{0.1}Y_{0.3}O_{2.8}$ (s) | –1597.30 ± 5.02 | This work |

Here: solution 1 is a 1 M solution of HCl with 0.1 M KI.